\documentclass[preprint,12pt]{elsarticle}

\def\lbldef#1#2{\expandafter\gdef\csname #1\endcsname {#2}}

\def\href#1#2{#2}

\usepackage{color}
\usepackage{graphicx}
\usepackage{amssymb}
\journal{Physics of the Dark Universe}

\begin{document}
	
\begin{frontmatter}
		
\title{Strong Observational Support for the $R_{\rm h}=ct$ \\ Timeline in the Early Universe}
		
\author[1]{Fulvio Melia\footnote{John Woodruff Simpson
	Fellow. E-mail: fmelia@email.arizona.edu}} 
		
\address[1]{Department of Physics, The Applied Math Program, and Department of Astronomy,
	The University of Arizona, AZ 85721, USA}
		
\begin{abstract}
{\it JWST}'s recent discovery of well-formed galaxies and supermassive black holes
only a few hundred Myr after the big bang seriously challenges the timeline predicted by $\Lambda$CDM.
Now, the latest identification of polycyclic aromatic hydrocarbons (PAHs) at $z=6.71$,
together with these earlier inconsistencies, makes the time compression problem in this
model quite overwhelming. We consider the timeline associated with the formation and growth
of PAH grains based on current astrophysical models and argue that their appearance at $z=6.71$
favors the structure formation history in $R_{\rm h}=ct$ rather than that of {\it Planck}-$\Lambda$CDM.
We estimate the time at which they must have started growing in each case, and then trace their
history through various critical events, such as the end of the `dark ages,' the beginning of Pop
III star formation, and the onset of reionization. Together, these three distinct 
discoveries by {\it JWST}, viz. high-$z$ galaxies, high-$z$ quasars and the surprisingly early
appearance of PAHs, all paint a fully consistent picture in which the timeline in $\Lambda$CDM
is overly compressed at $z\gtrsim 6$, while strongly supporting the expansion history in the early
Universe predicted by $R_{\rm h}=ct$.
\end{abstract}
		
\end{frontmatter}
	
\section{Introduction}\label{intro}
The first set of observations of the early Universe by the James
Webb Space Telescope ({\it JWST}) has already produced some surprising
results challenging the timeline predicted by {\it Planck}-$\Lambda$CDM.
Well formed, billion solar-mass galaxies and supermassive black holes
have been discovered at redshifts corresponding to a mere few hundred
Myr after the big bang in this model, inconsistent with conventional
astrophysical models for the formation of the earliest (Pop III) stars
at $t\sim 300$ Myr and the subsequent assembly of massive structures.

Now there appears to be an independent third discovery of a particular type
of dust that should have taken over a billion years to form on the conventional
asymptotic branch (AGB) channel, yet is present within a galaxy at $z=6.71$,
corresponding to a mere $\sim 500$ Myr after the first stars appeared 
\cite{Witstok:2023}. Large dust reservoirs ($\sim 10^8\;M_\odot$)
had already been detected in galaxies beyond $z\sim 8$
\cite{Watson:2015,Tamura:2019,Witstok:2022}. Generating this amount of dust
only $\sim 300$ Myr after the first stars were created is already quite
challenging. But the discovery of a 2,175~$\AA$ attenuation feature,
due to polycyclic aromatic hydrocarbons (PAHs), much earlier than expected
in conventional models for the growth of such particles, has considerably
worsened the implied time compression problem for the standard model.

In this paper, we demonstrate that all three time compression problems
highlighted by {\it JWST} point to a common failure in the standard time versus
redshift relation, strongly suggesting that the problem lies---not with the
conventional astrophysics of galaxy formation, supermassive black hole
growth and the creation of dust in the interstellar medium, but---with the
presumed background cosmology itself. In contrast, all of the {\it JWST}
observations of the early Universe are fully consistent with the timeline
predicted by the alternative Friedmann-Lema\^itre-Robertson-Walker (FLRW)
cosmology known as the $R_{\rm h}=ct$ universe
\cite{Melia:2007,MeliaShevchuk:2012,Melia:2020}.

\section{Expansion History}\label{expansion}
Modern cosmology is based on the FLRW metric
\cite{Weinberg:1972,Melia:2020}. This well-known solution to Einstein's equations, however,
gives no guidance concerning the equation of state, $p=w\rho$, in the cosmic fluid. The
standard model partitions the total energy density $\rho$ (and total pressure
$p$) into three primary components \cite{Ostriker:1995}: matter $\rho_{\rm m}$, radiation
$\rho_{\rm r}$, and an unknown dark energy $\rho_{\rm de}$, which is often assumed to be a
cosmological constant, $\Lambda$, with $w_\Lambda=-1$.  For the other two components, the
standard model assumes that $w_{\rm m}=0$ and $w_{\rm r}=1/3$.

The $R_{\rm h}=ct$ universe is also an FLRW cosmology, with a similar partitioning of
energy components in the cosmic fluid, but goes one step further by honoring the zero
active mass condition from general relativity, constraining the total equation of
state to be $\rho+3p=0$ \cite{Melia:2007,MeliaShevchuk:2012,Melia:2020}. This 
condition appears to be required for the self-consistent use of the FLRW metric
\cite{Melia:2022b,Melia:2023}.

\begin{figure}
\centering
\includegraphics[width=4.8in]{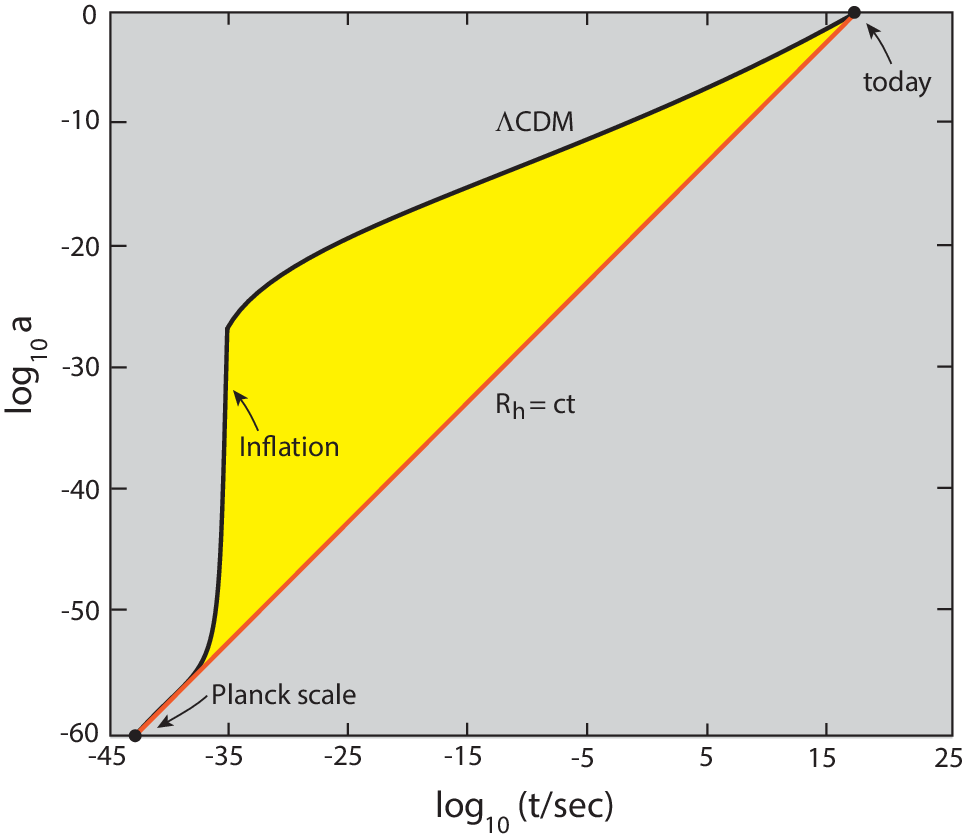}
\caption{Expansion history of the Universe in {\it Planck}-$\Lambda$CDM (solid black
curve), compared to the corresponding rate in the $R_{\rm h}=ct$ universe (red line).
Though the latter model has no horizon problems and does not need the poorly understood
inflationary scenario, the standard model cannot correctly account for the appearance of
the Universe today without a spurt of de Sitter-like expansion at $t\sim 10^{-37}$ s.
Yet in spite of various phases of deceleration and acceleration, the overall expansion
of the $\Lambda$CDM universe over a Hubble time is identical to what it would have
been from over 60 magnitudes of steady expansion in $R_{\rm h}=ct$. This equality has
essentially zero probability of occurring now, just when we happen to be looking, and
constitutes the most serious coincidence in standard cosmology.}
\label{fig1}
\end{figure}

Model selection based on more than 30 different kinds of data have already demonstrated
that $R_{\rm h}=ct$ is favored over $\Lambda$CDM with a likelihood of $\gtrsim 90\%$ versus
$\lesssim 10\%$ for the standard model (see, e.g., Table 2 in ref.~\cite{Melia:2018e}).
The majority of this work has been
focused on observations at $z\lesssim 8$, however, where the predictions of $\Lambda$CDM often
come close to those of $R_{\rm h}=ct$. In figure~\ref{fig1}, this region of overlap occupies
the very righthand side of the plot, at $t\gtrsim 10^{15}$ s. These curves,
based on Equation~(\ref{eq:tLambda}) for $t\gtrsim 10^{-37}$ s and $a(t)\propto
e^{H_{\rm inf}t}$ during inflation for $\Lambda$CDM (with $H_{\rm inf}$ the constant
Hubble parameter in de Sitter), and correspondingly on Equation~(\ref{eq:tRh}) for 
$R_{\rm h}=ct$, reveal several intriguing linkages between the two models. For example, 
given that $\Lambda$CDM is characterized by a larger number of free parameters than 
$R_{\rm h}=ct$, one may reasonably infer from this comparison that the optimized 
$\Lambda$CDM distance and expansion rate are merely mimicking those of the more 
fundamentally motivated model with $w=-1/3$.

This question was partially addressed in \cite{Melia:2015a}, where the zero active mass
condition was shown to provide an explanation for the optimized value $\sim 0.3$ of the
fraction $\Omega_{\rm m} \equiv \rho_m/\rho$ in $\Lambda$CDM---an otherwise seemingly
random number. This result appears to be a direct consequence of trying to fit the data
with the equation of state $w=(\rho_{\rm r}/3-\rho_{\rm de})/\rho$ in a Universe whose
principal constraint is instead $w=-1/3$, as expected in $R_{\rm h}=ct$ \cite{Melia:2022b}.

But the message we glean from the profiles in figure~\ref{fig1} is much more profound
than this, and constitutes the main goal of this paper. This plot clearly shows
that, while one may be able to mimic $R_{\rm h}=ct$ with $\Lambda$CDM at $z\lesssim 8$
via an appropriate choice of parameter values, this approach fails progressively more
and more as we examine the expansion history towards earlier times. For example, while
$R_{\rm h}=ct$ has no temperature or electroweak horizon problems
\cite{Melia:2013c,Melia:2018e}, $\Lambda$CDM cannot survive without inflation at
$t\sim 10^{-37}$ seconds in order to explain the Universe we see around us today
\cite{Starobinskii:1979,Kazanas:1980,Guth:1981,Linde:1982}.

Even so, the complicated history of acceleration and deceleration applied to the
standard model produces an overall expansion today {\it exactly equal} to what it would
have been in $R_{\rm h}=ct$ anyway. This is arguably the most inexplicable coincidence in
standard cosmology, i.e., that in spite of the infinite range of expansion profiles permitted
by the parameterization in $\Lambda$CDM, our Universe today has an apparent (i.e., Hubble)
horizon \cite{Melia:2018b} $R_{\rm h}$ exactly equal to $ct$ (hence the eponymous origin
of the name ``$R_{\rm h}=ct$"). The probability of this occurrence happening right now,
when we just happen to be looking, is effectively zero, because it could never have
happened earlier or at any time in our (presumably) infinite future.

\section{{\it JWST} Observations}\label{observations}
These are among the reasons why the most recent observations by {\it JWST} 
have provided an unprecedented challenge to $\Lambda$CDM,
by allowing us to probe the formation of large scale structure a mere $\sim 200$
Myr after the big bang---according to the timeline in this model. In just two
years, we have acquired high precision data pertaining to three kinds of source,
all of which---as we shall show in this paper---consistently demonstrate
that the timeline in the standard model (to the left of $t\sim 10^{15}$ s in
figure~\ref{fig1}) is strongly disfavored by the observations.

Two of these, the premature formation of (i) well-formed galaxies \cite{Melia:2023b} and 
(ii) $10^9\;M_\odot$ supermassive black holes \cite{Melia:2024b}, have already been 
discussed elsewhere, but we shall briefly summarize the key points shortly for completeness. 
Both the high-$z$ galaxies ($z\sim 16$) and quasars ($z\sim 10$) independently argue strongly
in favor of the timeline in $R_{\rm h}=ct$. Together they offer compelling observational
support for this model. But when the latest discovery reported by {\it JWST} just this year
is added to this body of evidence, the case against $\Lambda$CDM appears to be
overwhelming.

As we shall see, Witstok et al. \cite{Witstok:2023} found strong evidence of the ultraviolet 
attenuation `bump' (first discovered in the Milky Way by Stecher et al. \cite{Stecher:1965}) 
attributed to PAHs, i.e., nano-sized graphitic grains \cite{Li:2001}, in the spectrum of a 
galaxy at $z=6.71$. Basic astrophysical principles would require a much longer time for 
these carbonaceous dust grains to form than the age of the $\Lambda$CDM universe at 
that redshift (see figure~\ref{fig2}). In contrast, all three of these categories of 
source self-consistently follow the timeline predicted by the $R_{\rm h}=ct$ universe 
(figure~\ref{fig3}).

Throughout this paper, we determine the age-redshift relation in 
$\Lambda$CDM using the expression 
\begin{equation}
        t^{\Lambda{\rm CDM}}(z) = {1\over H_0}\int_z^\infty {du\over
        \sqrt{\Omega_{\rm m}(1+u)^3+\Omega_{\rm r}(1+u)^4+\Omega_\Lambda}}\label{eq:tLambda}
\end{equation}
where, in keeping with the basic assumption of a flat Universe and dark energy in the form
of a cosmological constant, we have simply $\Omega_\Lambda=1-\Omega_{\rm m}-\Omega_{\rm r}$.
The corresponding expression for $R_{\rm h}=ct$ is
\begin{equation}
        t^{R_{\rm h}}(z) = {1\over H_0(1+z)}\;.\label{eq:tRh}
\end{equation}
As demonstrated elsewhere \cite{MeliaFatuzzo:2016}, these relations yield the same age
for the Universe today in both models (see fig.~\ref{fig1}), but $t^{R_{\rm h}}$
is roughly twice $t^{\Lambda{\rm CDM}}$ at $z\gtrsim 6$, where the expansion
factor is $a\sim 1/7$ (with $a_0=1$). This is the reason why the $R_{\rm h}=ct$ 
universe easily eliminates the `too early' galaxy and supermassive black 
hole problems, as we shall see.

\subsection{Early Galaxies}\label{galaxies}
A large number of high-$z$ galaxy candidates ($z> 12$) was discovered by {\it JWST}
in less than a month of operation. Some were identified through the Early Release Observations
(ERO) \cite{Pontoppidan:2022}, others via the Cosmic Evolution Early Release Science (CEERS)
\cite{Finkelstein:2022} and Through the Looking GLASS (GLASS-{\it JWST}) \cite{Treu:2022}
programs. At least up to $z\sim 13$, the galaxies' photometric redshift has already been
confirmed spectroscopically \cite{Robertson:2022}.

The challenge to $\Lambda$CDM is that well-formed $\sim 10^9\,M_\odot$ stellar aggregates
at $z\sim 16-17$ would have emerged only $\sim 230$ Myr after the big bang \cite{Melia:2014a,Melia:2023b}.
This conflicts with the expected formation of structure in the standard model, based on
the {\it Planck} optimized parameters: a Hubble constant, $H_0=67.4 \pm0.5$ km s$^{-1}$
Mpc$^{-1}$, a matter density $\Omega_{\rm m}=0.315\pm0.007$, scaled to today's critical
density ($\equiv 3c^2H_0^2/ 8\pi G$), and a spatial curvature constant, $k\approx 0$
\cite{PlanckVI:2020}.

Their early appearance raises at least two questions: (i) Was the gas budget in the
early Universe sufficient to account for this galaxy demographic \cite{Behroozi:2018}?
The answer could be yes as long as all of the baryons in the host halos condensed into
stars \cite{Donnan:2022}; (ii) Can the dynamics of structure account for the anomalously
rapid formation of these galaxies
\cite{Yajima:2022,Keller:2022,Kannan:2022,Inayoshi:2022,Haslbauer:2022,Mirocha:2023,Whitler:2023}?
The availability of gas is one thing, but it is the dynamics and cooling rate of the plasma
that govern how soon after the big bang stars could have condensed and formed billion solar-mass
structures.

And this is the problem, because none of the simulations thus far have produced a billion
new stars in only $\sim 70-90$ Myr by the time the Universe was only $\sim 230$ Myr old. 
Our current view of how the earliest galaxies formed is based on a broad range
of simulations developed over the past several decades to trace the growth of initial 
fluctuations seen in the cosmic microwave background. The first suite of calculations 
\citep{Barkana:2001,Miralda:2003,Bromm:2004,Ciardi:2005,Glover:2005,Greif:2007,Wise:2008,Salvaterra:2011,Greif:2012,Jaacks:2012}
elucidated how Pop~III stars formed by redshift $z\sim 20$ within dark-matter halos with 
mass $M_{\rm halo}\sim 10^6\,M_\odot$ \citep{Haiman:1996,Tegmark:1997,Abel:2002,Bromm:2002}. 
This delay after the big bang does not appear to be surmountable due to the influence of 
several processes, including the initial gravitational growth of the dark-matter perturbations 
and, especially, the inefficient, radiative cooling of the primordial gas. As far as we know
today, the condensation into stars followed the slow cooling via molecular hydrogen 
\citep{Galli:1998,Omukai:1998}. In $\Lambda$CDM, the Universe was $\sim 180$ Myr old at 
redshift 20. More recent simulations have improved this constraint somewhat by showing
that halos could have been distributed across this age 
\citep{Yajima:2022,Keller:2022,Kannan:2022,Inayoshi:2022,Haslbauer:2022,Mirocha:2023,Whitler:2023},
some appearing as early as $\sim 120$ Myr after the big bang, though still too late to
alleviate the time delay.

The baryonic gas clouds condensing within these halos formed protostars at their
center, growing to become $> 100\,M_\odot$ Pop~III stars \citep{Kroupa:2002,Chabrier:2003}. 
But the UV radiation from such massive stars destroyed all of the $H_2$ that had formed, 
limiting each halo to only a few Pop~III stars \citep{Yoshida:2008}. 

Then, the reheating and expulsion of the surrounding gas from these first stars 
further delayed the formation of new stars until the fluctuations in the early
Universe had cooled and condensed again to high densities. This process would have
taken another $\sim 100$ Myr---roughly the dynamical time for a first-galaxy halo to 
assemble \citep{Yoshida:2004,Johnson:2007}. By this time, the $\Lambda$CDM Universe was
$\sim 300$ Myr old.

\begin{figure}
\centering
\includegraphics[width=4.8in]{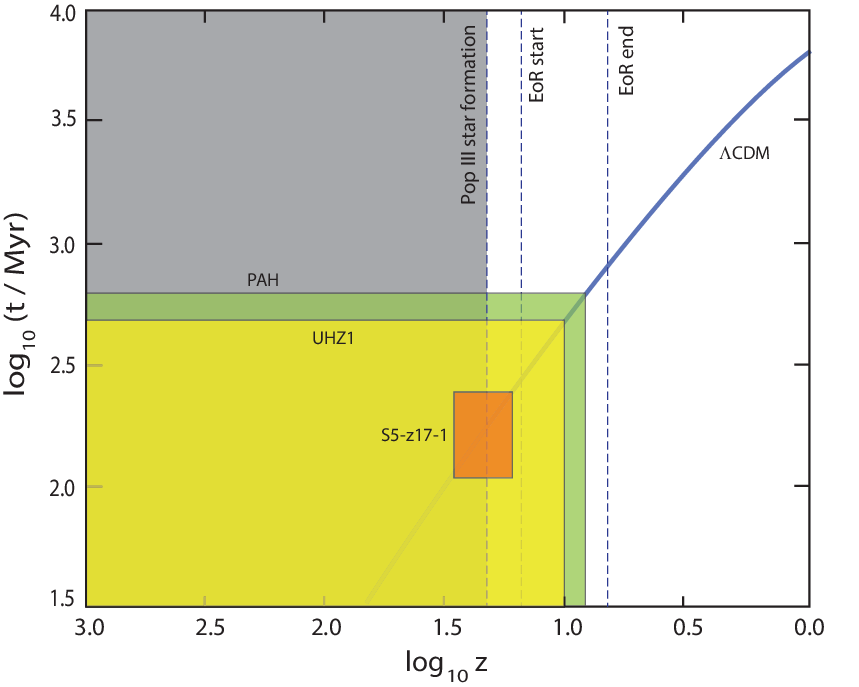}
\caption{Cosmic time versus redshift in {\it Planck}-$\Lambda$CDM (solid blue curve).
In this model, the `dark ages' ended after Pop III stars started forming no earlier than
$z\sim 20$, corresponding to $t\sim 200$ Myr. Observationally, the EoR started at 
$z\sim 15$ ($t\sim 280$ Myr) and ended at $z\sim 6$ ($t\sim 1$ Gyr).
With standard astrophysics, the earliest known quasar (UHZ-1) would have needed to start
forming before the big bang ($t<0$), and the earliest galaxies (e.g., S5-z17-1) would
have been seeded before any stars could have appeared. The PAHs discovered at $z=6.71$
($t\sim 900$ Myr) should have taken $\sim 1$ Gyr to form, also starting before the big
bang, and certainly well before the first Pop III stars were created.}
\label{fig2}
\end{figure}

\begin{figure}
\centering
\includegraphics[width=4.8in]{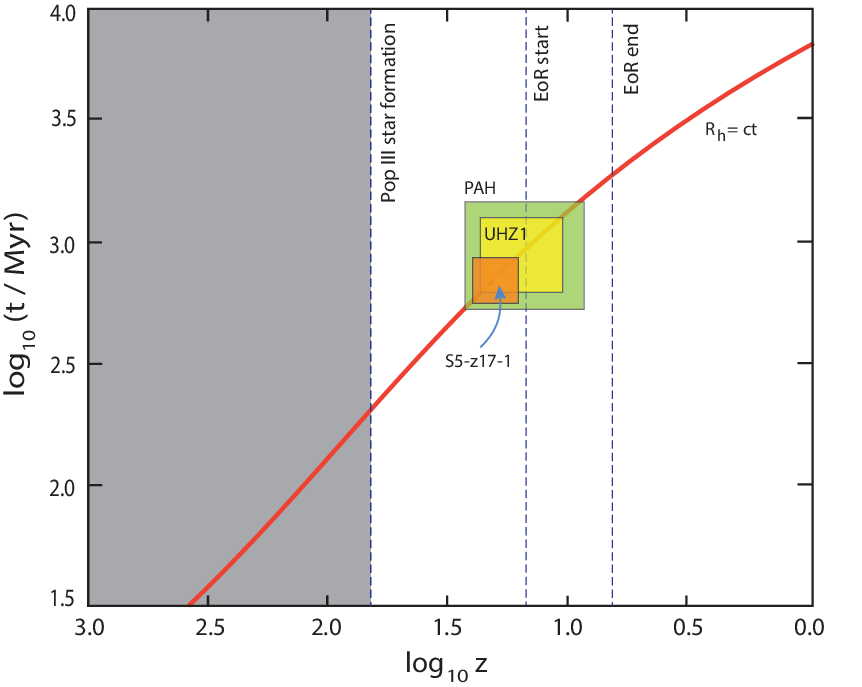}
\caption{Same as figure~\ref{fig2}, except here for $R_{\rm h}=ct$ (solid red curve).
In this model, the `dark ages' would have ended at $z\sim 50$ ($t\sim 250$ Myr) and
the EoR would have started at $t\sim 820$ Myr ($z\sim 15$) and ended at $\sim 1.9$ Gyr
($z\sim 6$). The PAHs appear at $\sim 1.7$ Gyr ($z=6.71$), after having formed since
$z\sim 17$ ($t\sim 730$ Myr). UHZ-1 was seeded at $z\sim 20$ ($t\sim 620$ Myr) and we
now see it at $t\sim 1.2$ Gyr ($z=10.1$). The galaxy S5-z17-1 initiated its growth at
$t\sim 500$ Myr ($z\sim 25$) and we see it at $\sim 750$ Myr ($z\sim 16$).}
\label{fig3}
\end{figure}

Roughly speaking, this picture suggests that the $\sim 10^9$ stars in a galaxy
such as S5-z17-1 (fig.~\ref{fig2}) must have formed in only $\sim 70-90$ Myr,
but none of the simulations to date have been able to explain this, even with
the introduction of innovative features and physical processes designed to 
reduce the inconsistencies as much as possible
\citep{Yajima:2022,Keller:2022,Kannan:2022,Inayoshi:2022,Haslbauer:2022,Mirocha:2023,Whitler:2023}.
For example, introducing a large scatter in the cooling times \cite{Yajima:2022,Keller:2022}
and weaker Pop~III supernovae that expel less gas \citep{Kitayama:2005,Frebel:2015} 
reduces the tension somewhat, but still not enough to account for the highest-$z$ galaxies
observed thus far. 

The simulations of Jaacks et al. \cite{Jaacks:2012} are not necessarily the most
up-to-date available now, but they are easy to interpret and understand how the factors 
discussed above come into play \cite{Melia:2023b}. These investigators examined the duty 
cycle and history of the star-formation rate at $z\gtrsim 6$ and showed that it can be 
represented well by an exponentially increasing function with characteristic timescale
$t_c\sim 70+33(\log_{10}M-6)$ Myr, for galaxies with stellar mass $M\sim 10^6\;M_\odot$ to
$\sim 10^{10}\;M_\odot$. From their simulations, one may therefore infer a growth rate
for these early galaxies given by the expression
\begin{equation}
	{dM\over dt}=K\exp\left({t-t_*\over t_c}\right)\;,\label{eq:Jaacks}
\end{equation}
where $t_*$ is the time at which the galaxy first formed and $K$ is the average 
star-formation rate which, in the case of S5-z17-1, is $\sim 9.7\;M_\odot$ yr$^{-1}$
\cite{Harikane:2022}.

Based on this empirical formula, one thus concludes that S5-z17-1 would
have started forming at $t_*\sim 107$ Myr, which corresponds to $z_*\sim 28$ in
$\Lambda$CDM. In figure~\ref{fig2}, the seeding and growth of this galaxy is
represented by the red rectangle, whose lower left corner sits at $107$ Myr
and its upper right corner is at $z=16.66$.

Clearly, the problem with this timeline is that the hot gas had not even cooled and 
condensed yet at $t_*$ \cite{Melia:2023b}. In other words, Pop III stars had just 
barely started forming, so there could not have been any billion solar-mass galaxies 
completed by that time.

This time compression problem completely disappears for the $R_{\rm h}=ct$ cosmology.
In this model, the expansion factor scales as $a(t)\propto t$, yielding the simple age-redshift
relation $1+z={t_0/t}$, calibrated to the current age, $t_0$ of the Universe. Since the 
gravitational radius $R_{\rm h}$ coincides with the Hubble radius, we have $t_0={1/ H_0}$.
The corresponding initiation and growth of S5-z17-1 in the context of
$R_{\rm h}=ct$ is represented by the red box in figure~\ref{fig3}. As one can see in 
this plot, the epoch of reionization (EoR) redshift range $6< z< 15$ corresponds to 
$906\;{\rm Myr}< t < 2.07$ Gyr in this model. The Universe would thus have had ample 
time to construct billion-solar mass galaxies after Pop~III and Pop~II stars began 
condensing in numbers. Notice especially that all of the high-$z$ galaxies discovered 
thus far appear after the dark ages ended, as one should expect if they contributed 
or dominated the subsequent reionization \cite{Munoz:2024,Melia:2024c}. To demonstrate 
an age-redshift inconsistency in $R_{\rm h}=ct$ like that seen in $\Lambda$CDM, one would 
need to discover a billion-solar mass galaxy at $z\sim 50$.

\subsection{Early Quasars}\label{quasars}
The time compression problem in the early $\Lambda$CDM universe has been emphasized more
than ever by the recent discovery \cite{Bogdan:2023} of the X-ray luminous supermassive
black hole, UHZ-1, at a redshift $z=10.073\pm 002$ \cite{Goulding:2023}. 

In conventional astrophysics, black-hole growth (regardless of its mass) 
is constrained by the maximum accretion rate permitted by the outward radiation pressure
due to the luminosity emitted via the dissipation of gravitational energy \citep{Melia:2009}. 
For hydrogen-rich plasma, this limit is characterized by the Eddington value,
$L_{\rm Edd}\approx 1.3\times 10^{38}(M/M_\odot)$ ergs s$^{-1}$. The efficiency, $\epsilon$, 
representes the rate at which rest-mass energy is converted into radiation, such that 
$\dot{M}=L_{\rm bol}/\epsilon c^2$, with $L_{\rm bol}$ the bolometric luminosity, 
which is often different from $L_{\rm Edd}$. The general consensus is that
$\epsilon\sim 0.1$ when all possible variations of accretion-disk theory are
taken into account. 

When the accretion rate is Eddington-limited, one may thus write
\begin{equation}
{dM\over dt}={1.3\times 10^{38}\;{\rm ergs/s}\over\epsilon c^2M_\odot}\;M\label{eq:dMdt}
\end{equation}
\citep{Salpeter:1964,Melia:2013a}, whose solution, 
\begin{equation}
M(t) = M_{\rm seed}\exp\left({t-t_{\rm seed}\over 45\;{\rm Myr}}\right),\label{eq:Salpeter}
\end{equation}
is commonly referred to as the Salpeter relation. In this expression, $M_{\rm seed}$ is
the seed mass created at time $t_{\rm seed}$. For supernova remnants, one typically has
$M_{\rm seed}\sim 5-25\;M_\odot$.

The illustrative case, UHZ-1, shown in figures~\ref{fig2} and \ref{fig3} has an
inferred mass $M=10^7-10^8\;M_\odot$ which, according to Equation~(\ref{eq:Salpeter}), 
should have taken over $\sim 600$ Myr to grow via standard Eddington-limited accretion 
starting with a mass of $\sim 10\;M_\odot$, if one adopts the supernova-remnant seeding
typically assumed for such objects \cite{Melia:2020,Melia:2024b}.  

But in standard cosmology, the age of the Universe at $z=10.073$ was only $466$ Myr,
which one may argue was $\lesssim 300$ Myr after Pop III stars started forming. This
redshift corresponds to the upper righthand corner of the yellow boxes in figures~\ref{fig2}
and \ref{fig3}. As one can clearly see in the case of $\Lambda$CDM, not only would UHZ-1
have started growing before the earliest stars appeared but, worse, it would have been
seeded even before the big bang, if the early quasars such as this evolved according to
standard astrophysical principles. In contrast, UHZ-1 would have been created at 
$z\sim 20$ in the context of $R_{\rm h}=ct$, when the Universe was $\sim 620$ Myr old,
and we see it $\sim 600$ Myr later at $z=10.073$. The yellow box in figure~\ref{fig3}
is thus fully contained within the period following the formation of Pop~III stars,
allowing ample time for them to evolve and die as supernovae to create the necessary
$\gtrsim 10\;M_\odot$ seeds. Moreover, all three sources considered in this paper
coincide with the start of the EoR in this model, whereas they precede the transition
from the dark ages to the onset of reionization in $\Lambda$CDM, which is difficult
to reconcile with the general understanding of what produced the ionizing photons
in the first place \cite{Munoz:2024,Melia:2024c}.

Attempts at resolving this inconsistency have focused on two speculative scenarios:
(i) greatly super-Eddington accretion \cite{Volonteri:2005,Pacucci:2015,Inayoshi:2016} and
(ii) an exotic formation of $\sim 10^5\;M_\odot$ seeds via direct collapse
\cite{Yoo:2004,Latif:2013,Alexander:2014}. But no evidence has ever been found
of super-Eddington accretion, either in UHZ-1 or any other supermassive black hole.
Quite generally, the inferred luminosity in all high-$z$ quasars with reasonably
estimated masses is near the Eddington value (see, e.g., fig.~5 in
ref.~\cite{Willott:2010a}). For example, the bolometric luminosity of J0313-1806 ($z=7.642$)
is $0.67\pm0.14\;L_{\rm Edd}$ \cite{Wang:2021b}, that of J1342+0928 ($z=7.54$) is
$1.5^{+0.5}_{-0.4}\;L_{\rm Edd}$ \cite{Banados:2018}, and J1007+2115 ($z=7.515$) is
at $1.06\pm0.2\;L_{\rm Edd}$ \cite{Yang:2020}.

The massive seed scenario is even more challenging to confirm observationally. The
dynamic events creating such objects would be too brief for direct detection.
Once formed, they could be found nearby, but no conclusive identification
has yet been made. The creation of massive seeds is contemplated theoretically,
but their existence is highly speculative.

In contrast to the significant tension created for standard cosmology by the early
emergence of these supermassive black holes, all of their characteristics,
including their mass and redshift, supplemented by our current models of Pop III
star formation and death as supernovae, are fully consistent with the timeline
in $R_{\rm h}=ct$ (see figure~\ref{fig3}) \cite{Melia:2013b,MeliaMcClintock:2015,
Fatuzzo:2017}. As was the case for the `too early' appearance of well formed
galaxies, this time compression problem faced by $\Lambda$CDM is completely
eliminated in $R_{\rm h}=ct$.

Each of these categories of source, high-$z$ galaxies and supermassive black holes,
individually creates significant tension for the time versus redshift relation
predicted by {\it Planck}-$\Lambda$CDM. Together, they refute the standard
timeline rather strongly. And when we next consider the discovery of
PAHs far too early in the history of the $\Lambda$CDM universe---a third
{\it JWST} `strike' against structure formation in this model---the combined,
self-consistent argument of all three against the expansion rate expected in
$\Lambda$CDM appears to be overwhelming.

\subsection{PAHs in the Early Universe}\label{PAH}
Deep Near-Infrared Spectrograph (NIRSpec) multi-object spectroscopic observations
with {\it JWST} in the spectral range $0.6$ to $5.3\;\mu$m and resolving power
$R\approx 100$ have revealed strong evidence of an absorption feature near the
rest frame wavelength $\lambda_{\rm emit}=2175\;\AA$ in the spectrum of the galaxy
JADES-GS+53.15138-27.81917 (JADES-GS-z6-0 for short) \cite{Witstok:2023}. This feature
is commonly referred to as the ultraviolet attenuation bump, produced by carbonaceous
dust grains, specifically polycyclic aromatic hydrocarbons (PAHs). These are
nano-sized graphitic grains \cite{Li:2001} seen at $z=6.71$, corresponding to
an age $t\lesssim 890$ Myr in the {\it Planck}-$\Lambda$CDM universe, and may
have formed even earlier, based on the detection of large dust reservoirs (up
to $\sim 10^8\;M_\odot$) in galaxies out to $z\sim 8$, when the
{\it Planck}-$\Lambda$CDM universe was only $\sim 600$ Myr old
\cite{Watson:2015,Tamura:2019,Witstok:2022}.

The detection of PAHs so early in the Universe's history does not comport very
well with our understanding of how these grains formed. Beyond our local neighborhood,
this feature has been observed previously only in massive, metal-enriched galaxies
at $z\lesssim 3$, implying it originates in dust grains created solely in evolved
galaxies \cite{Shivaei:2022,Noll:2009}. Indeed, a measurement of the gas-phase and
stellar metallicity ($Z\sim 0.2-0.3\;Z_\odot$) in JADES-GS-z6-0 suggests that it
experienced substantial metal enrichment relative to other galaxies seen with a similar
mass and redshift \cite{Curti:2023}. In Witstok et al.'s (2023) analysis, e.g., no
bump was seen in a sample of galaxies ($4\lesssim z\lesssim 11.5$) with emission from
the C III$\lambda$1,907, 1,909 $\AA$ nebular lines commonly seen in metal-poor galaxies.

The galaxy JADES-GS-z6-0 thus appears to be much older than the age implied by
an interpretation of the observations using the $\Lambda$CDM timeline. The conventional
picture of how carbon and the carbonaceous grains producing the absorption feature at
$z=6.71$ formed relies on the standard AGB channel, especially
at low metallicties. The detection of PAHs at this redshift corresponds to a time in
$\Lambda$CDM when the oldest stars were only $\sim 500$ Myr old. But low to intermediate
mass stars ($\sim 0.5$ to $\sim 8.0\;M_\odot$) would not have been old enough by then
to have produced dust in their AGB phase. In the Milky Way, stardust production
is dominated by AGB stars, which typically take at least $\sim 1$ Gyr (see
figure~\ref{fig2}) to reach this stage \cite{Seok:2014,Li:2020a,Yang:2023}.

The top righthand corner of the green box in figures~\ref{fig2} and 
\ref{fig3} corresponds to the redshift ($z=6.71$) at which JADES-GS-z6-0 was
discovered, which represents an age $t\sim 900$ Myr in $\Lambda$CDM and
$\sim 1.7$ Gyr in $R_{\rm h}=ct$. Based on the time required for typical AGB stars 
to produce the observed PAHs as seen in all of the available simulations completed
thus far \cite{Seok:2014,Li:2020a,Yang:2023}, we estimate the time at which
JADES-GS-z6-0 was created to have been at least $\sim 1$ Gyr earlier than this.
The green box thus extends to a time before the big bang in $\Lambda$CDM, but only
to $t\sim 730$ Myr in $R_{\rm h}=ct$, which corresponds to a redshift $z\sim 17$
in this model. We thus see the same pattern as described above for the
earliest galaxies and supermassive black holes, i.e., that all of these sources
must have formed well before the earliest stars appeared on the scene---and actually
even before the big bang---in the standard model, but typically around the redshift 
($z\sim 20$) coincident with, or following, the transition from the dark ages to
the onset of the EoR in $R_{\rm h}=ct$.

If the traditional explanation for the formation of PAHs in JADES-GS-z6-0 is abandoned
for being too slow and inefficient, alternative scenarios for their growth include
their formation via more massive and rapidly evolving stars, such as supernovae and
Wolf-Rayet (WR) stars. This possibility is motivated by the fact that only stars with
masses $\gtrsim 8\;M_\odot$ could have evolved in under $\sim 500$ Myr.

But supernovae type-Ib/c explosions generally destroy most dust produced in the preceding
WR phase, and carbon-rich WR stars are rare \cite{Eldridge:2017}. Substantial carbonaceous
production in SN ejecta may occur only in some classes of models, e.g., non-rotating
progenitors \cite{Kirchschlager:2019}. Thus, any attempt at explaining the formation of
PAHs much more rapidly than is possible via the AGB channel is quite speculative and
poorly motivated based on the models and simulations available today.

As things stand today, it is quite remarkable that the $\Lambda$CDM timeline 
appears to fail in very similar ways for all three classes of source we have considered 
in this paper. The time compression problem would be solved in every case if the sources 
were created several hundred Myr before the big bang, a similarity that is itself unavoidably 
suspicious. This `coherence' is amplified by the fact that all three classes of source fit 
neatly and self-consistently within a tiny region of the $t-z$ phase space for the timeline 
in $R_{\rm h}=ct$ (figure~\ref{fig3}), consistent with standard astrophysical models for the 
growth of structure. Their observed characteristics are entirely in line with the earliest
formation of (Pop III) stars at $t\sim 300$ Myr, followed by the subsequent
assembly of billion solar mass galaxies, the supernova deaths of these (and Pop II)
stars and the growth of supermassive black holes over the next $\sim 1$ Gyr, all the
way to $z\sim 16$ (galaxies), $\sim 10$ (quasars) and $\sim 6.7$ (PAHs), at which we
detect these sources.

\section{Conclusion}\label{conclusion}
As one can see in figure~\ref{fig2} (green shaded box), the formation of PAHs 
in JADES-GS-z6-0 is inconsistent with the timeline predicted by $\Lambda$CDM. The AGB stars 
producing these nano-sized graphitic grains would have had to be over $\sim 1$ Gyr old by then,
meaning that they must have been born several hundred Myr before the big bang, echoing the
time compression problem created by the high-$z$ galaxies and supermassive black holes
discussed in \S\S~\ref{galaxies} and \ref{quasars}.

The principal caveat to this work is that alternative mechanisms for producing PAHs
in the early Universe may mitigate the time compression problem implied by the appearance
of these complex molecules containing two or more carbon-based rings so early compared
to the time required in their conventional AGB branch channel. Possibilities other than
the supernova and WR star production discussed above include their creation in strong
interstellar shocks \cite{Tielens:1987}, ion-molecule reactions in dense inerstellar
clouds \cite{Herbst:1991}, and the accretion of C$+$ ions \cite{Omont:1986}. The
onus, of course, is on proponents of the standard model to show quantitatively how
these methods work in the context of $\Lambda$CDM. Unfortunately, it has not been
possible to place any observational constraints on these alternative PAS formation
methods thus far.

In this paper, however, we have argued that the time compression problem represented by
the latest {\it JWST} observations is not isolated to just some inconsequential category
of source. Strong tension with the timeline predicted by $\Lambda$CDM is created by all
of these data, including the too-early appearance of well formed billion solar-mass galaxies,
supermassive black holes and PAHs. And quite tellingly, while speculative, exotic mechanisms
are being introduced to preserve the time versus redshift relation in the standard model, all
of the time compression problems disappear completely and self-consistently in the
context of $R_{\rm h}=ct$.

\section*{Acknowledgements}
I am grateful to the anonymous referee for their comments and suggestions.

\bibliographystyle{elsarticle-harv}
\bibliography{ms}

\end{document}